# ELASTIC PROTON SCATTERING ON $^{13}$C AND $^{15}$C NUCLEI IN THE DIFFRACTION THEORY

## M.A. Zhusupov[1], E.T. Ibraeva[2], R.S. Kabatayeva[1]

[1] - Al-Farabi Kazakh National University, [2] - Institute of Nuclear Physics
Almaty, Kazakhstan

**ABSTRACT.** There is a calculation of the differential cross sections of proton scattering on $^{13}$C and $^{15}$C nuclei at energy of 1 GeV with the shell model wave functions in the framework of the Glauber theory in the study. The single, double and triple collisions have been taken into account in the multiple scattering operator. The role of each term of the series and their interference in the differential cross section have been estimated. It is shown that for a description of the cross sections in a wide angle/momentum transfer range it is necessary to consider not only the first, but the higher scattering orders.

## INTRODUCTION

A study of non-stable, neutron- or proton-rich isotopes put before the nuclear physics the fundamental questions of determination of the nucleon stability boundaries, nuclear shells evolution, synthesis of super-heavy elements in accelerators and cosmic objects. The phenomena unknown until now are observed: inhomogeneity of neutron and proton densities (halo), new deformations areas and new types of collective excitations at low energies (soft dipole resonance), breakings in nuclear shells occupation, appearing of new magic numbers and other. The carbon isotopes are intensively studied what is concerned with both the wide distribution of the stable $^{12}$C isotope and the obtaining of beams of non-stable $^{15,16,17,19,20,22}$C isotopes at both the average (from 40 to 103 MeV per nucleon [1 − 5]) and the high (500 – 740 MeV per nucleon [6, 7]) energies at the facilities in GANIL, NSCL MSU, RIKEN and other. Recently the carbon targets have been used for getting of tensor-polarized deuterons as a result of the effect called the nuclear spin dichroism. The first experiment was carried out in Cologne at the tandem accelerator which uses the carbon target bombarded by deuterons [8]. In Saint-Petersburg Institute for Nuclear Physics in Gatchina and in the Nuclear Centre in Sakley the differential cross sections were measured at energy E = 1 GeV for the number of nuclei $^9$Be, $^{10,11}$B, $^{12,13}$C and other in the angle range $25° > \theta > 5°$ [9]. The calculations of the differential cross sections were carried out in the framework of the Glauber diffraction scattering theory [10] which reproduces them well for the forward angles (until the second diffraction minimum ($\theta \sim 20°$)).

At the fragment-separator FRS in GSI one obtained the beams of $^{12-19}$C isotopes and the cross sections of scattering of protons and $^{12}$C isotope on these isotopes at the energy of E = 950 MeV per nucleon were measured for the first time. The cross sections with the proton target show the sharp growth with an increase of the neutrons number in the projectile particle [11].

In paper [12] the dilatational momentum distributions of the $^{13,14}$C fragments from the $^{15}$C and the $^{13}$C fragments from the $^{14}$C were measured at 83 MeV per nucleon at the cyclotron in Riken Projectile Fragment Separator in RIKEN Ring Cyclotron Facility. An analysis of these momentum distributions and the cross sections of reactions of one- or two-neutrons stripping was carried out in the Glauber model basing on the core plus neutron model.

An analysis of the cross sections of $^{12-22}$C + p reaction at energies from 40 to 800 MeV in the framework of the Glauber theory was carried out in the paper [5]. The wave functions of the carbon isotopes are generated on the base of the simple model of average field. The evaluation of contributions of the neutron and proton cross sections into the elastic p + C - scattering shows that the nuclear surface gives the main contribution into the cross section ($\sigma_R$) what is especially noticed at the low energies (40, 100 MeV) [5]. In the elastic p + $^{15}$C - scattering when energy changes from 40 to 800 MeV the value of $\sigma_R$ varies from 580 to 294 mb. Discussing the advantages and disadvantages of the proton and carbon targets the authors [5] concluded that the proton target could probe the surface of the neutron-rich nuclei better than the $^{12}$C target, especially at low energies.

The prove for the extended structure of the ground and the first excited states of $^{15}$C nucleus is presented in paper [13] when studying the elastic magnet electron scattering on the $^{15,17,19}$C isotopes. It is shown that one can determine the orbital of the last nucleon from the elastic magnet form-factor. The form-factor at the large values in the momenta space is defined by the density at the small distances in the coordinate space and vice versa. The form-factors at the low and intermediate momentum transfers for all the $^{15,17,19}$C isotopes are similar to each other when the last neutron occupies the same orbital (let admit that $2s_{1/2}$). This means that the asymptotics of the wave functions defined by the last neutron are also close to each other. At the same time there is an explicit difference between the form-factors for the same nucleus when the last neutron occupies the different orbitals ($2s_{1/2}$ or $1d_{5/2}$).

The total reaction cross sections of carbon isotopes with N = 6 – 16 in collisions with the proton targets in the energy range 40 – 1000 MeV were systematically studied [5]. The core plus neutron model was chosen for the odd-nucleon nuclei. The empirical formulas were suggested which are used for the predictions of unknown total reaction cross sections. A parametrization of the NN-amplitude of scattering with one and two gaussoids are discussed. It is shown that the standard parametrization of the amplitude by one gaussoid gives nearly the same numerical results for the cross sections of the reaction that the two gaussoids do that is why it is enough to use one.

In paper [14] the pair correlations in the odd isotopes of carbon and the influence of Pauli principle account in interactions like particle plus core in the $^{13}$C and $^{11}$Be nuclei are considered. In the two approaches BCS (Bardin-Cuper-Shiff) and PBCS (projectile BCS) the energy spectra and single-particle spectroscopic factors of the low-lying states of $^{13,15,17,19}$C isotopes were calculated and a comparison with experimental data was carried out. The results show that the pair interaction plays an important role in the nuclear structure of the heavy carbon isotopes due to their characteristics: the small binding energy and the systematic decrease of the binding energy at adding of neutrons (to 5 MeV) when transiting from the $^{13}$C to the $^{19}$C.

The stable 3α - linear chain structure of the $^{13}$C isotopes is studied in the microscopic cluster 3α + n - model [15-17]. Two excited rotational series were found, where three α-cluster structures of the $^{13}$C isotope were obtained [15]. The structure of 1/2$^{\pm}$ states in the $^{13}$C near the 3α + n - threshold ($E_x$ = 12.3 MeV) are studied in the complete 3α + n orthogonal states model (OCM) [16]. In addition to the 3α + n - model the effect of decay of one of the α-clusters is taken into account. The nuclear densities of the low-lying states are calculated [17].

In the shell model the structures of the ground states of the C$^{13}$ and C$^{15}$ nuclei are different [18]. If the ground state of the C$^{13}$ nucleus has the shell configuration | (1s)$^4$ (1p)$^9$: J$^\pi$, T = 1/2$^-$, 1/2 >, then the C$^{15}$ nucleus in the ground state is | (1s)$^4$ (1p)$^{10}$ 2s$_{1/2}$: J$^\pi$, T = 1/2$^+$, 3/2 >. The structure of the C$^{15}$ nucleus in the ground state is close to the one of the excited state in the C$^{13}$ nucleus with quantum numbers J$^\pi$ = 1/2$^+$ at E ~ 4 MeV and has the shell structure | (1s)$^4$ (1p)$^8$ 2s$_{1/2}$: J$^\pi$, T = 1/2$^+$, 1/2 >.

The objective of the present study is the "microscopic" calculation and the analysis of the differential cross sections of elastic scattering of protons on $^{13}$C and $^{15}$C nuclei in the framework of the Glauber theory with the wave functions in the shell model at energy of 1 GeV, and also a study of the interference effects in the Glauber multiple scattering operator.

## BRIEF FORMALISM

In the Glauber diffraction scattering theory the matrix elements of p$^{13}$C - scattering are calculated at energy of 1 GeV. The analogous calculation for the p$^{15}$C - scattering was carried out in our paper [19]. The input parameters of the theory are the wave functions and the parameters of the proton-nucleon amplitude extracted usually from independent experiments.

The wave function in the shell model for the $^{13}$C nucleus is presented by the shell configuration $|(1s)^4(1p)^9\rangle$ in the ground state [18].

Let us represent the shell wave function in a form

$$\Psi_{i,f}(\vec{r}_1,...\vec{r}_{13}) = |(1s)^4(1p)^9\rangle = \Psi_{n_0 l_0 m_0}(\vec{r}_1,...\vec{r}_4)\Psi_{n_1 l_1 m_1}(\vec{r}_5,...\vec{r}_{13}), \qquad (1)$$

where $n_i l_i m_i$ − quantum numbers of the corresponding shell ($n_0 = 0, l_0 = 0, m_0 = 0$; $n_1 = 1, l_1 = 1, m_1 = 0, \pm 1$). Each function is a product of single-particle one

$$\Psi_{nlm}(\vec{r}_1,\vec{r}_2,...) = \prod_{\nu} \Psi_{nlm}(\vec{r}_\nu).$$

The matrix element (amplitude)

$$M_{if}(\vec{q}) = \frac{ik}{2\pi}\int d^2\vec{\rho}\exp(i\vec{q}\vec{\rho})\langle\Psi_f^{JM_J}|\Omega|\Psi_i^{JM_J'}\rangle, \qquad (2)$$

here $\vec{q}$ − is a momentum transfer in a reaction:

$$\vec{q} = \vec{k} - \vec{k}', \qquad (3)$$

$\vec{k},\vec{k}'$ − incoming and outgoing momenta of the incident and escape proton. In case of elastic scattering $|\vec{k}| = |\vec{k}'|$ and the momentum $q$ equals:

$$q = 2k\sin\frac{\theta}{2}, \quad k = \sqrt{\varepsilon^2 - m^2}, \qquad (4)$$

The operator of multiple scattering:

$$\Omega = 1 - \prod_{\nu=1}^{13}\left(1-\omega_\nu(\vec{\rho}-\vec{\rho}_\nu)\right) = \sum_{\nu=1}^{13}\omega_\nu - \sum_{\nu\langle\tau=1}^{13}\omega_\nu\omega_\tau + \sum_{\nu\langle\tau\langle\eta=1}^{13}\omega_\nu\omega_\tau\omega_\eta - ...(-1)^{12}\omega_1\omega_2...\omega_{13}. \qquad (5)$$

Limiting to triple collisions, substituting the operator into the formula (2) and having integrated it with respect to $d\vec{\rho}$, $d\vec{q}_i,...,d\vec{q}_k$, one obtains

$$\tilde{\Omega} = \tilde{\Omega}^{(1)} - \tilde{\Omega}^{(2)} + \tilde{\Omega}^{(3)} = $$
$$= \frac{2\pi}{ik}f_{pN}(q)\sum_{\nu=1}^{13}\tilde{\omega}_\nu - \left(\frac{2\pi}{ik}f_{pN}\left(\frac{q}{2}\right)\right)^2\sum_{\nu<\tau=1}^{13}\tilde{\omega}_\nu\tilde{\omega}_\tau + \left(\frac{2\pi}{ik}f_{pN}\left(\frac{q}{3}\right)\right)^3\sum_{\nu<\tau<\eta=1}^{13}\tilde{\omega}_\nu\tilde{\omega}_\tau\tilde{\omega}_\eta - ... \qquad (6)$$

The «tilde» sign over the operators means the integrations done.

$$\sum_{\nu=1}^{13}\tilde{\omega}_\nu = \sum_{\nu=1}^{13}\exp(i\vec{q}\vec{\rho}_\nu), \qquad (7)$$

$$\sum_{\nu<\tau=1}^{13}\tilde{\omega}_\nu\tilde{\omega}_\tau = \sum_{\nu<\tau=1}^{13}\exp\left(i\frac{\vec{q}}{2}(\vec{\rho}_\nu+\vec{\rho}_\tau)\right)\delta(\vec{\rho}_\nu-\vec{\rho}_\tau), \qquad (8)$$

$$\sum_{\nu<\tau<\eta=1}^{13}\tilde{\omega}_\nu\tilde{\omega}_\tau\tilde{\omega}_\eta = \sum_{\nu<\tau=1}^{13}\exp\left(i\frac{\vec{q}}{3}(\vec{\rho}_\nu+\vec{\rho}_\tau+\vec{\rho}_\eta)\right)\delta(\vec{\rho}_\nu-\vec{\rho}_\tau)\delta\left(\vec{\rho}_\eta-\frac{1}{2}(\vec{\rho}_\nu+\vec{\rho}_\tau)\right). \qquad (9)$$

The matrix element (2) in the approach of triple scattering:

$$M_{if}(\vec{q}) = M_{if}^{(1)}(\vec{q}) - M_{if}^{(2)}(\vec{q}) + M_{if}^{(3)}(\vec{q}), \qquad (10)$$

where

$$M_{if}^{(1)}(\vec{q}) = \frac{ik}{2\pi}\left\langle \Psi_f \left| \sum_{\nu=1}^{13} \tilde{\omega}_\nu \right| \Psi_i \right\rangle, \tag{11}$$

$$M_{if}^{(2)}(\vec{q}) = \frac{ik}{2\pi}\left\langle \Psi_f \left| \sum_{\nu<\tau=1}^{13} \tilde{\omega}_\nu \tilde{\omega}_\tau \right| \Psi_i \right\rangle, \tag{12}$$

$$M_{if}^{(3)}(\vec{q}) = \frac{ik}{2\pi}\left\langle \Psi_f \left| \sum_{\nu<\tau<\eta=1}^{13} \tilde{\omega}_\nu \tilde{\omega}_\tau \tilde{\omega}_\eta \right| \Psi_i \right\rangle. \tag{13}$$

The matrix element of single scattering:

$$M_{if}^{(1)}(\vec{q}) = \frac{ik}{2\pi}\left\langle \prod_{\nu=1}^{4}\Psi_{000}(\vec{r}_\nu)\prod_{\tau=5}^{13}\Psi_{11m}(\vec{r}_\tau) \left| \sum_{\nu=1}^{4}\tilde{\omega}_\nu + \sum_{\tau=5}^{13}\tilde{\omega}_\tau \right| \prod_{\nu=1}^{4}\Psi_{000}(\vec{r}_\nu)\prod_{\tau=5}^{13}\Psi_{11m}(\vec{r}_\tau) \right\rangle \tag{14}$$

$$M_{if}^{(1)}(\vec{q}) = 4\, N_{11} M_{if}^{(1)-s}(\vec{q}) + 9\, N_{00} M_{if}^{(1)-p}(\vec{q}), \tag{15}$$

where

$$M_{if}^{(1)-s}(\vec{q}) = \frac{ik}{2\pi}\left\langle \prod_{\nu=1}^{4}\Psi_{000}(\vec{r}_\nu)\left|\sum_{\nu=1}^{4}\tilde{\omega}_\nu\right|\prod_{\nu=1}^{4}\Psi_{000}(\vec{r}_\nu)\right\rangle = \frac{ik}{2\pi}\int \prod_{\nu=1}^{4}|\Psi_{000}(\vec{r}_\nu)|^2 \sum_{\nu=1}^{4}\tilde{\omega}_\nu\, d\vec{r}_\nu, \tag{16}$$

$$M_{if}^{(1)-p}(\vec{q}) = \frac{ik}{2\pi}\left\langle \prod_{\nu=5}^{13}\Psi_{11m}(\vec{r}_\nu)\left|\sum_{\nu=5}^{13}\tilde{\omega}_\nu\right|\prod_{\nu=5}^{13}\Psi_{11m}(\vec{r}_\nu)\right\rangle = \frac{ik}{2\pi}\int \prod_{\nu=5}^{13}|\Psi_{11m}(\vec{r}_\nu)|^2 \sum_{\nu=5}^{13}\tilde{\omega}_\nu\, d\vec{r}_\nu, \tag{17}$$

$$N_{00} = \left\langle \prod_{\nu=1}^{4}\Psi_{000}(\vec{r}_\nu) \middle| \prod_{\nu=1}^{4}\Psi_{000}(\vec{r}_\nu) \right\rangle, \quad N_{11} = \left\langle \prod_{\nu=5}^{13}\Psi_{11m}(\vec{r}_\nu) \middle| \prod_{\nu=5}^{13}\Psi_{11m'}(\vec{r}_\nu) \right\rangle. \tag{18}$$

$M_{if}^{(1)-s}(\vec{q})$ − is a matrix elements of scattering on nucleons of 1s-shell (4 terms), $M_{if}^{(1)-p}(\vec{q})$ − on nucleons of 1p-shell (9 terms), $N_{00}$, $N_{11}$ − normalization of wave functions.

A calculation of these matrix elements with the wave functions in the shell model in the spherical coordinates system with change of the two-dimensional $\vec{\rho}$ vector for the three-dimensional $\vec{r}$ vector and an expansion of the expression $\exp(i\vec{q}\vec{\rho}_\nu)$ in a series by Bessel $J_{\lambda+\frac{1}{2}}(qr)$ functions and spherical $Y_{\lambda\mu}(\Omega_q)$ harmonics give one the following result:

$$M_{if}^{(1)-s}(\vec{q}) = 4 N_{00}^{3/4}\sqrt{\frac{\pi}{2q}}\int_0^\infty |R_{00}(r)|^2 J_{\frac{1}{2}}(qr) r^{3/2} dr, \tag{19}$$

$$M^{(1)-p}(\vec{q}) = 10(N_{11m})^{9/10}\sqrt{\frac{\pi}{2q}}\sqrt{4\pi}\sum_{\lambda\mu m}(i)^\lambda\sqrt{2\lambda+1}\, B_{11}^{(\lambda)}(q)\langle\lambda 010|10\rangle\langle\lambda\mu 1m'|1m\rangle Y_{\lambda\mu}(\Omega_q), \tag{20}$$

where

$$B_{11}^{(\lambda)}(q) = \int_0^\infty |R_{11}(r)|^2 J_{\lambda+1/2}(qr) r^{5/2} dr \ . \tag{21}$$

The radial wave functions

$$R_{00} = C_{00} \exp\left(-\frac{r^2}{2r_0^2}\right), \quad C_{00} = \frac{2}{\pi^{1/4} r_0^{3/2}}, \tag{22}$$

$$R_{11} = C_{11} \frac{r}{r_0} \exp\left(-\frac{r^2}{2r_0^2}\right), \quad C_{11} = \sqrt{\frac{2}{3}} C_{00} \ . \tag{23}$$

Here $r_0$ is connected with the oscillatory parameter $\hbar\omega$ by the relation $r_0^2 = \frac{\hbar}{m\omega} = \frac{(\hbar c)^2}{mc^2 \hbar\omega} = \frac{(1.97 \cdot 10^{-11} \text{MeV} \cdot \text{cm})^2}{940 \text{MeV} \cdot 14 \text{MeV}}$. For the 1p-shell nuclei $\hbar\omega = 13.8$ MeV [20].

The matrix element of double scattering

$$M_{if}^{(2)}(\vec{q}) = \frac{ik}{2\pi} \Bigg\langle \prod_{\nu=1}^{4} \Psi_{000}(\vec{r}_\nu) \prod_{\tau=5}^{13} \Psi_{11m}(\vec{r}_\tau) \Bigg| \sum_{\nu<\tau=1}^{4} \tilde{\omega}_\nu \tilde{\omega}_\tau + \\
+ \sum_{\nu=1}^{4} \tilde{\omega}_\nu \sum_{\tau=5}^{13} \tilde{\omega}_\tau + \sum_{\nu=5}^{12} \tilde{\omega}_\nu \sum_{\tau=6}^{13} \tilde{\omega}_\tau \Bigg| \prod_{\nu=1}^{4} \Psi_{000}(\vec{r}_\nu) \prod_{\tau=5}^{13} \Psi_{11m'}(\vec{r}_\tau) \Bigg\rangle \tag{24}$$

$$M_{if}^{(2)}(\vec{q}) = 6 M_{000}^{(2)-ss}(\vec{q}) N_{11} + 36 M_{0011}^{(2)-sp}(\vec{q}) + 36 M_{11}^{(2)-pp}(\vec{q}) N_{00}, \tag{25}$$

$$M_{if}^{(2)-ss}(\vec{q}) = \frac{2\pi}{k} f_{pN}^2\left(\frac{q}{2}\right) \Bigg\langle \prod_{\nu=1}^{4} \Psi_{000}(\vec{r}_\nu) \Bigg| \sum_{\nu<\tau=1}^{4} \tilde{\omega}_\nu \tilde{\omega}_\tau \Bigg| \prod_{\nu=1}^{4} \Psi_{000}(\vec{r}_\nu) \Bigg\rangle = \\
= \frac{2\pi}{k} f_{pN}^2\left(\frac{q}{2}\right) \int \prod_{\nu=1}^{4} |\Psi_{000}(\vec{r}_\nu)|^2 \sum_{\nu<\tau=1}^{4} \exp\left(i\frac{\vec{q}}{2}(\vec{r}_\nu + \vec{r}_\tau)\right) \delta(\vec{r}_\nu - \vec{r}_\tau) \prod_{\nu=1}^{4} d\vec{r}_\nu, \tag{26}$$

$$M_{if}^{(2)-sp}(\vec{q}) = \frac{2\pi}{k} f_{pN}^2\left(\frac{q}{2}\right) \Bigg\langle \prod_{\nu=1}^{4} \Psi_{000}(\vec{r}_\nu) \Bigg| \sum_{\nu=1}^{4} \tilde{\omega}_\nu \Bigg| \prod_{\nu=1}^{4} \Psi_{000}(\vec{r}_\nu) \Bigg\rangle \Bigg\langle \prod_{\nu=5}^{13} \Psi_{11m}(\vec{r}_\nu) \Bigg| \sum_{\nu=5}^{13} \tilde{\omega}_\nu \Bigg| \prod_{\nu=5}^{13} \Psi_{11m'}(\vec{r}_\nu) \Bigg\rangle \tag{27}$$

$$M_{if}^{(2)-pp}(\vec{q}) = \frac{2\pi}{k} f_{pN}^2\left(\frac{q}{2}\right) \Bigg\langle \prod_{\nu=5}^{13} \Psi_{11m}(\vec{r}_\nu) \Bigg| \sum_{\nu<\tau=5}^{13} \tilde{\omega}_\nu \tilde{\omega}_\tau \Bigg| \prod_{\nu=5}^{13} \Psi_{11m'}(\vec{r}_\nu) \Bigg\rangle = \\
= \frac{2\pi}{k} f_{pN}^2\left(\frac{q}{2}\right) \int \prod_{\nu=5}^{13} |\Psi_{11m}(\vec{r}_\nu)|^2 \sum_{\nu<\tau=5}^{13} \exp\left(i\frac{\vec{q}}{2}(\vec{r}_\nu + \vec{r}_\tau)\right) \delta(\vec{r}_\nu - \vec{r}_\tau) \prod_{\nu=5}^{13} d\vec{r}_\nu \tag{28}$$

where $M_{if}^{(2)-ss}(\vec{q})$, $M_{if}^{(2)-sp}(\vec{q})$, $M_{if}^{(2)-pp}(\vec{q})$ − matrix elements of collisions of proton with the nucleon on (1s)-, (1s, 1p)-, (1p)-shells. Let us denote $C^{(2)} = \frac{2\pi}{k} f_{pN}^2\left(\frac{q}{2}\right)$.

Applying the same technique as at calculating the single scattering matrix element one gets finally

$$M_{if}^{(2)-ss}(\vec{q}) = 4\frac{\sqrt{\pi}}{\sqrt{2q}}C^{(2)}\int_0^\infty |R_{00}(r)|^2 J_{1/2}(qr)r^{3/2}dr. \tag{29}$$

$$M_{if}^{(2)-pp}(\vec{q}) = 36\frac{\sqrt{\pi}}{\sqrt{2q}}C^{(2)}\sum_{\lambda\mu m} B_{11}^{(\lambda)}(q)F_{\lambda\mu}(\Omega_q), \tag{30}$$

where

$$B_{11}^{(\lambda)}(q) = \int_0^\infty |R_{11}(r)|^2 J_{\lambda+1/2}(qr)r^{5/2}dr \tag{31}$$

$$\begin{aligned}F_{\lambda\mu}(\Omega_q) &= \int Y_{1m}(\Omega_r)Y_{1m}^*(\Omega_r)Y_{\lambda\mu}(\Omega_r)d\Omega_r Y_{\lambda\mu}(\Omega_q) = \\ &= \sum_{\lambda\mu m}\sqrt{\frac{(2\lambda+1)}{4\pi}}\langle 10\lambda 0|10\rangle\langle 1m\lambda\mu|1m'\rangle Y_{\lambda\mu}(\Omega_q)\end{aligned} \tag{32}$$

The matrix element of triple scattering

$$M_{if}^{(3)}(\vec{q}) = \frac{ik}{2\pi}\left\langle \Psi_{000}(\vec{r}_1,...\vec{r}_4)\Psi_{11m_1}(\vec{r}_5,...\vec{r}_{13})\left|\tilde{\Omega}^{(3)}\right|\Psi_{000}(\vec{r}_1,...\vec{r}_4)\Psi_{11m_1}(\vec{r}_5,...\vec{r}_{13})\right\rangle\prod_{\nu=1}^A d\vec{r}_\nu \tag{33}$$

The operator of triple scattering

$$\begin{aligned}\tilde{\Omega}^{(3)} &= \sum_{\nu<\tau<\eta}\tilde{\omega}_\nu\tilde{\omega}_\tau\tilde{\omega}_\eta = \\ &= \sum_{\nu<\tau<\eta=1}^4 \tilde{\omega}_\nu\tilde{\omega}_\tau\tilde{\omega}_\eta + \sum_{\nu<\tau=1}^4 \tilde{\omega}_\nu\tilde{\omega}_\tau \sum_{\eta=5}^{13}\tilde{\omega}_\eta + \sum_{\nu=1}^4 \tilde{\omega}_\nu \sum_{\tau<\eta=5}^{13}\tilde{\omega}_\tau\tilde{\omega}_\eta + \sum_{\nu<\tau<\eta=5}^{13}\tilde{\omega}_\nu\tilde{\omega}_\tau\tilde{\omega}_\eta.\end{aligned} \tag{34}$$

Such a representation of the $\tilde{\Omega}^{(3)}$ operator allows one to calculate the contribution into the differential cross sections from re-scattering on the different shells

$$M_{if}^{(3)}(\vec{q}) = 4M_{if}^{(3)-sss}(\vec{q}) + 54M_{if}^{(3)-ssp}(\vec{q}) + 144M_{if}^{(3)-spp}(\vec{q}) + 84M_{if}^{(3)-ppp}(\vec{q}), \tag{35}$$

where

$$\begin{aligned}M_{if}^{(3)-sss}(\vec{q}) &= \left(\frac{2\pi}{ik_0}\right)^3 4f_{pN}^3\left(\frac{q}{3}\right)\left\langle\Psi_f\left|\sum_{\nu<\tau<\eta=1}^4\tilde{\omega}_\nu\tilde{\omega}_\tau\tilde{\omega}_\eta\right|\Psi_i\right\rangle = \left(\frac{2\pi}{ik_0}\right)^3 4f_{pN}^3\left(\frac{q}{3}\right)\times \\ &\times\left\langle\prod_{\nu=1}^4\Psi_{000}(\vec{r}_\nu)\left|\sum_{\nu<\tau<\eta=1}^4\tilde{\omega}_\nu\tilde{\omega}_\tau\tilde{\omega}_\eta\right|\prod_{\nu=1}^4\Psi_{000}(\vec{r}_\nu)\right\rangle\end{aligned} \tag{36}$$

$$\begin{aligned}M_{if}^{(3)-ssp}(\vec{q}) &= \left(\frac{2\pi}{ik_0}\right)^3 54f_{pN}^3\left(\frac{q}{3}\right)\left\langle\Psi_f\left|\sum_{\nu<\tau=1}^4\tilde{\omega}_\nu\tilde{\omega}_\tau\sum_{\eta=5}^{13}\tilde{\omega}_\eta\right|\Psi_i\right\rangle = \left(\frac{2\pi}{ik_0}\right)^3 54f_{pN}^3\left(\frac{q}{3}\right)\times \\ &\times\left\langle\prod_{\nu=1}^4\Psi_{000}(\vec{r}_\nu)\prod_{\nu=5}^{13}\Psi_{11m}(\vec{r}_\nu)\left|\sum_{\nu<\tau=1}^4\tilde{\omega}_\nu\tilde{\omega}_\tau\sum_{\eta=5}^{13}\tilde{\omega}_\eta\right|\prod_{\nu=1}^4\Psi_{000}(\vec{r}_\nu)\prod_{\nu=5}^{13}\Psi_{11m'}(\vec{r}_\nu)\right\rangle,\end{aligned} \tag{37}$$

$$M_{if}^{(3)-spp}(\vec{q}) = \left(\frac{2\pi}{ik_0}\right)^3 144 f_{pN}^3 \left(\frac{q}{3}\right) \left\langle \Psi_f \left| \sum_{\nu=1}^{4} \tilde{\omega}_\nu \sum_{\tau<\eta=5}^{13} \tilde{\omega}_\tau \tilde{\omega}_\eta \right| \Psi_i \right\rangle = \left(\frac{2\pi}{ik_0}\right)^3 144 f_{pN}^3 \left(\frac{q}{3}\right) \times$$

$$\times \left\langle \prod_{\nu=1}^{4} \Psi_{000}(\vec{r}_\nu) \prod_{\nu=5}^{13} \Psi_{11m}(\vec{r}_\nu) \left| \sum_{\nu=1}^{4} \tilde{\omega}_\nu \sum_{\tau<\eta=5}^{13} \tilde{\omega}_\tau \tilde{\omega}_\eta \right| \prod_{\nu=1}^{4} \Psi_{000}(\vec{r}_\nu) \prod_{\nu=5}^{13} \Psi_{11m'}(\vec{r}_\nu) \right\rangle, \quad (38)$$

$$M_{if}^{(3)-ppp}(\vec{q}) = \left(\frac{2\pi}{ik_0}\right)^3 84 f_{pN}^3 \left(\frac{q}{3}\right) \left\langle \Psi_f \left| \sum_{\nu<\tau<\eta=5}^{13} \tilde{\omega}_\nu \tilde{\omega}_\tau \tilde{\omega}_\eta \right| \Psi_i \right\rangle = \left(\frac{2\pi}{ik_0}\right)^3 84 f_{pN}^3 \left(\frac{q}{3}\right) \times$$

$$\times \left\langle \prod_{\nu=5}^{13} \Psi_{11m}(\vec{r}_\nu) \left| \sum_{\nu<\tau<\eta=5}^{13} \tilde{\omega}_\nu \tilde{\omega}_\tau \tilde{\omega}_\eta \right| \prod_{\nu=5}^{13} \Psi_{11m'}(\vec{r}_\nu) \right\rangle \quad (39)$$

The upper letter indices denote shells. The coefficients before the amplitudes show how many triple re-scatterings the proton undergoes with the nucleons from different shells.

After the variables separation and integration in the spherical coordinates system one gets the final expressions for the triple collisions matrix elements:

$$M_{if}^{(3)-sss}(\vec{q}) = D_1(q) \int \exp(-3r^2/r_0^2) J_{1/2}(qr) r^{3/2} dr, \quad (40)$$

$$D_1(q) = H(q) \frac{4 C_{00}^2}{2^2 \pi^{3/2}}, \quad H(q) = \left(\frac{2\pi}{ik_0}\right)^3 f_{pN}^3 \left(\frac{q}{3}\right) \frac{1}{\sqrt{2q}},$$

$$M_{if}^{(3)-ssp}(\vec{q}) = D_2(q) \sum_{\lambda \mu m m'} (i)^\lambda \sqrt{(2\lambda+1)} \langle \lambda 010|10\rangle \langle \lambda \mu 1m|1m'\rangle Y_{\lambda\mu}(\Omega_q) \times$$

$$\times \int \exp(-3r^2/r_0^2) J_{\lambda+1/2}(qr) r^{7/2} dr, \quad (41)$$

$$D_2(q) = H(q) \frac{54 C_{00}^2 C_{11}^2}{2^3 \pi r_0^2},$$

$$M_{if}^{(3)-spp}(\vec{q}) = D_3(q) \sum_{\lambda \mu m m'} \sum_{L M L' M'} (i)^\lambda (-1)^{m'} \sqrt{(2\lambda+1)} \langle 1010|L'0\rangle \langle L'010|L0\rangle \langle \lambda 0 L 0|10\rangle$$

$$\times \langle 1m_1 1m_2|L'M'\rangle \langle L'M'1-m_1'|LM\rangle \langle \lambda \mu LM|1m_1'\rangle Y_{\lambda\mu}(\Omega_q) \int \exp(-3r^2/r_0^2) J_{\lambda+1/2}(qr) r^{11/2} dr, \quad (42)$$

$$D_3(q) = H(q) \frac{144 C_{00}^2 C_{11}^2}{2 \pi r_0^4},$$

$$M_{if}^{(3)-ppp}(\vec{q}) = D_4(q) \sum_{\lambda \mu m m'} \sum_{L M L' M'} \sum_{K N K' N'} (i)^\lambda (-1)^{m_1'+m_2'+m_3'+N} \frac{\sqrt{(2\lambda+1)}}{(2K+1)} \langle 1010|L'0\rangle \langle L'010|L0\rangle \times$$

$$\times \langle 1010|K'0\rangle \langle K'010|K0\rangle \langle 1m_1 1m_2|L'M'\rangle \langle L'M'1-m_1'|LM\rangle \langle 1m_3'1-m_2'|K'N'\rangle \langle K'N'1-m_3'|KN\rangle \times$$

$$\times \langle \lambda 0 L 0|K 0\rangle \langle \lambda \mu LM|K-N\rangle Y_{\lambda\mu}(\Omega_q) \int \exp(-3r^2/r_0^2) J_{\lambda+1/2}(qr) r^{15/2} dr, \quad (43)$$

$$D_4(q) = H(q) \frac{84 \cdot 3^3 C_{11}^2}{2^3 \pi^2 r_0^6}.$$

$C_{00}$, $C_{11}$ – normalization coefficients of the radial wave functions, formulas (5), (6).

## RESULTS ANALYSIS

The calculation of the differential cross sections of protons elastic scattering on $^{13}C$ and $^{15}C$ nuclei at energy of 1 GeV in the Glauber theory has been done. Single-, double and triple-collisions have been taken into account in the operator of multiple scattering.

In figure 1 the differential cross sections of elastic $p^{13}C$-scattering at E = 1.0 GeV are shown with account of single- (dash-dot), double- (dashed), triple- (dots) collisions and the total cross section (solid curve). The experimental data are provided from the paper [9]. It is shown in the pictures, that at small angles the single scattering dominates, as the angle increases the single scattering decreases sharply, it is equated to the double scattering, and the last one defines the cross sections behavior at intermediate angles. With the further increase of the angle when the double collisions cross sections decrease, the triple collisions begin to play a role, the last ones describe the behavior of the cross sections. In the points of equality of the partial cross sections the minimums appear since the multiple scattering series is an alternating-sign one. At the energy of 1 GeV the dominating contribution of the single collisions is observed until 11° angles only, the double one – until 21°, the triple one – over 22°. All orders are present in the total differential cross section. The calculation describes the behavior of the differential cross sections correctly; some discrepancies are in the minimums since this area is more sensible to the structure of the wave function and to the interaction dynamics.

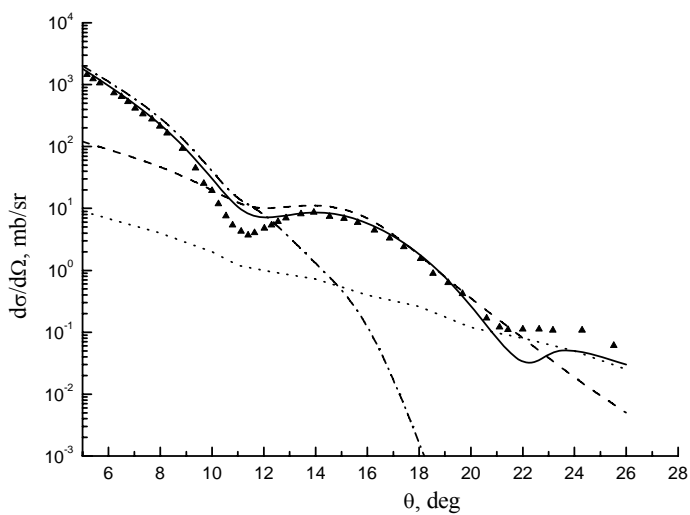

Figure 1. Contribution into the differential cross sections of $p^{13}C$-scattering: single- (dash-dot), double- (dash), triple- (dots) collisions and their sum (solid curve) at E = 1.0 GeV

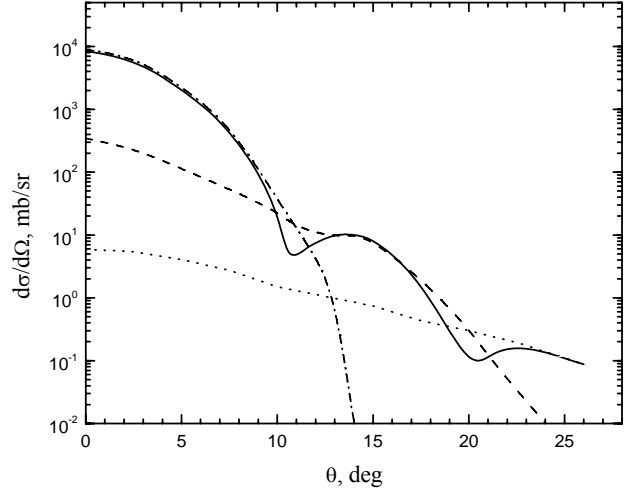

Figure 2. The same as in figure 1 for the $p^{15}C$-scattering

In figure 2 there is a contribution of the several first orders of scattering into the differential cross sections of $p^{15}C$-scattering. Each subsequent term of the series (5) gives a contribution of one order less than the previous one. As it was multiply pointed out in the literature (beginning from the [10]), the terms of the multiple scattering series decrease non-uniformly with the scattering angle increase: the single scattering decreases too sharply and even in the range of $\theta \sim 14°$ angles its contribution into the differential cross sections becomes negligibly small; as for the double scattering, being far less than the single one at small $\theta < 10°$ angles, it begins to dominate in the range of $20° > \theta > 12°$ angles and gives one the main contribution into the differential cross sections; the triple scattering term which is for two orders less than the single one at zero angle, is equated to the double one at $\theta \sim 20°$ angle and gives one the main contribution into the cross section at $\theta > 20°$. In the range where the curves of one-, double- and triple collisions intersect (i.e. where the differential cross sections are equated by absolute values), the interference minimums appear since the series (5) is an alternating-sign one and when squaring the matrix element the cross terms are subtracted. As it is seen from the result of our calculation in order to describe the behavior of the differential cross section in the wide angle range it is necessary to account not only single, but the higher orders of scattering either. This very result demonstrates enough fast convergence of the multiple scattering series (at the given energy of 1 GeV per nucleon), when it is enough to limit to the triple scattering since in the range of small angles (where the calculation by the Glauber theory is relevant) the orders higher than the third one will not give noticeable contribution into the cross section.

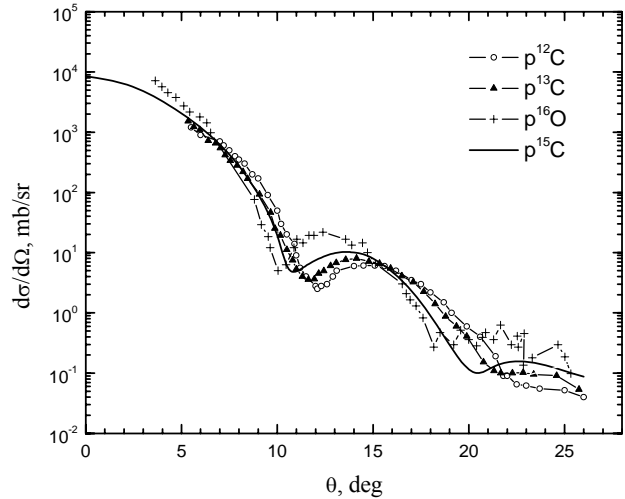

Figure 3. Comparison of the differential cross sections of the $^{12}$C, $^{13}$C, $^{15}$C, $^{16}$O nuclei at energy E=1,0 GeV per nucleon. Experimental data for $^{12}$C, $^{13}$C are from [21], for $^{16}$O − from [22]

In figure 3 there is a demonstration of the differential cross sections on $^{12}$C, $^{13}$C, $^{15}$C, $^{16}$O nuclei at energy E = 1.0 GeV. A comparison of the cross sections shows that with the growth of the nucleon number the cross section increases slightly at zero scattering angle, what reflects the growth of the mean-square radius (from $R_{rms}$ = 2.32 fm for $^{12}$C, till $R_{rms}$ = 3.42 fm for $^{16}$O) and the minimum in the cross section is shifted in the range of smaller scattering angles. The location and the value of the minimums differ significantly since the picture by the ordinate axis is in the logarithmic scale. The result obtained by us for the $^{15}$C nucleus does not contradict to the experimental data obtained for other carbon isotopes and for the $^{16}$O nucleus.

## CONCLUSIONS

There is a calculation technique of the proton elastic scattering matrix element on the $^{13}$C nucleus in the framework of the Glauber diffraction theory in the present study. A neglect of the "small" nuclear momenta $\vec{Q}_i$ in comparison with the transferred one $\vec{q}$, taken in the calculation, allowed us to calculate the entire series of multiple scattering, and the use of the shell wave functions in the harmonic oscillator basis provided analytical calculation of the dynamic integrals (dependent on the scattering operator in the arms of the wave functions). The obtained calculation formula of the matrix element presents a product of two multipliers, one of which is a sum of the multiple scattering series, and the other one stands for the scattering on nucleons from different shells of $^{15}$C nucleus. Thus it became possible to calculate the contribution into the differential cross sections from both the different orders of scattering and the scatterings on nucleons on the different shells.

Having calculated the contribution into the differential cross section from the first three terms of the multiple scattering series, we showed that the single collisions dominate at the very small scattering angles, until $\theta < 12°$, the double ones – in the range of the second maximum in the cross section at $20° > \theta > 12°$, the triple ones – in the range of the third maximum at $\theta > 20°$. In the intersection points of different scattering orders the typical interference minimums appear in the differential cross sections since the multiple scattering series is an alternating-sign one and the cross terms are subtracted from the total sum.

An absence of the experimental data for the p$^{15}$C-scattering motivated us to carry out a comparison of the differential cross sections of the close carbon $^{12}$C and $^{13}$C isotopes and the $^{16}$O nuclei at energy of 1 GeV, for the last one data are available. The cross section for the p$^{15}$C-scattering follows the typical peculiarities of others, being consistent with them by both the absolute value and the location of the minimums and the maximums.